\documentclass[aps,pra,groupedaddress,showpacs,twocolumn,superscriptaddress]{revtex4}
 
\usepackage{amsmath}
\usepackage{amssymb}
\usepackage{amsthm}

\newcommand{\ket}[1]{|#1\rangle}               
\newcommand{\bra}[1]{\langle #1|}              
\newcommand{\dyad}[2]{\ket{#1}\bra{#2}}        
\newcommand{\ip}[2]{\langle #1|#2\rangle}      
\newcommand{\matl}[3]{\langle #1|#2|#3\rangle} 

\newcommand{\HC}{\mathcal{H}}

\newcommand{\Tr}{\mathrm{Tr}}

\newtheorem{theorem}{Theorem}
\newtheorem{lemma}{Lemma}

\begin{document}

\title{Separable operations on pure states}

\author{Vlad Gheorghiu}
\email[Electronic address: ]{vgheorgh@andrew.cmu.edu}

\author{Robert B. Griffiths}
\affiliation{Department of Physics, Carnegie Mellon University, Pittsburgh,
Pennsylvania 15213, USA}

\date{Version of September 1, 2008}

\begin{abstract}
  We show that the possible ensembles produced when a separable operation
  acts on a single pure bipartite entangled state are completely characterized
  by a majorization condition, a collection of inequalities for Schmidt
  coefficients, which is identical to that already known for the particular
  case of local operations and classical communication (LOCC).  As a
  consequence, various known results for LOCC, including some involving
  monotonicity of entanglement, can be extended to the class of all separable
  operations.
\end{abstract}

\maketitle

\section{Introduction\label{sct1}} A separable operation $\Lambda$ on a
bipartite quantum system is a transformation of the form
\begin{equation}
\label{eqn1} \rho'=\Lambda(\rho)=\sum_{k=1}^N(A_k\otimes B_k)\rho(A_k\otimes
B_k)^\dagger,
\end{equation} where $\rho$ is an initial density operator on the Hilbert
space $\HC_A\otimes \HC_B$. The Kraus operators $A_k\otimes B_k$ are arbitrary
product operators satisfying the closure condition
\begin{equation}
\label{eqn2} \sum_{k=1}^NA_k^\dagger A_k\otimes B_k^\dagger B_k=I_A\otimes
I_B,
\end{equation} with $I_A$ and $I_B$ the identity operators. The extension to
multipartite systems is obvious, but here we will only consider the bipartite
case. To avoid technical issues the sums in \eqref{eqn1} and \eqref{eqn2} as
well as the dimensions $D_A$ and $D_B$ of $\HC_A$ and $\HC_B$ are assumed to
be finite.

Local operations with classical communication (LOCC) form a subset of
separable operations in which the Kraus operators $A_k\otimes B_k$ are
restricted by the requirement that they be generated in the following fashion.
Alice carries out an operation $\{A^{(1)}_i\}$, $\sum_i A^{(1)\dagger}_i
A^{(1)}_i=I_A$, in the usual way with the help of an ancilla, the measurement
of which yields the value of $i$, which is then transmitted to Bob.  He uses
$i$ to choose an operation $\{B^{(2,i)}_j\}$, the result $j$ of which is
transmitted back to Alice, whose next operation can depend on $j$ as well as
$i$, and so forth.  While it is (fairly) easy to see that the end result after
an arbitrary number of rounds is of the form \eqref{eqn1}, it is difficult to
characterize in simple mathematical or physical terms precisely what it is
that distinguishes LOCC from more general separable operations.  Examples show
that separable operations can be more effective than LOCC in distinguishing
certain sets of orthogonal states \cite{PhysRevA.59.1070}, even in a system as
simple as two qubits \cite{quantph.0705.0795}, but apart from this little is
known about the difference.

What we demonstrate in Sec.~\ref{sct2} of this paper is that the ensemble
$\{p_k,\ket{\phi_k}\}$ produced by a separable operation acting on a pure
state $\ket{\psi}$, see \eqref{eqn5}, satisfies a majorization condition
\eqref{eqn7}, which is already known to be a necessary and sufficient
condition for producing the same ensemble from the same $\ket{\psi}$ by LOCC.
Among the consequences discussed in Sec.~\ref{sct3} are: a separable
operation acting on a pure state can be ``simulated'' by LOCC; a necessary
condition for a deterministic transformation $\ket{\psi}\rightarrow\ket{\phi}$
given in \cite{PhysRevA.76.032310} can be replaced by a necessary and
sufficient majorization condition; and certain entanglement measures are
nonincreasing under separable operations.  Section~\ref{sct4} summarizes our
main result and indicates some open questions.

\section{Ensembles produced by separable operations on pure bipartite
states}\label{sct2}
\subsection{Majorization conditions\label{sct2A}}

Let $\{A_k\otimes B_k\}_{k=1}^{N}$ be a separable operation on
$\HC_A\otimes\HC_B$, specified by $N$ Kraus operators satisfying the closure
condition \eqref{eqn2}.  Let $\ket{\psi}$ be a normalized entangled state on
$\HC_A\otimes\HC_B$ with Schmidt form
\begin{equation}
\label{eqn3} \ket{\psi}=\sum_{j=1}^{D}\sqrt{\lambda_j}\ket{a_j}\ket{b_j},
\end{equation} where $D=D_B$, and we assume without loss of generality that
$D_A\geq D_B$.  Here $\{\ket{a_j}\}$ and $\{\ket{b_j}\}$ are orthonormal bases
chosen so that the Schmidt weights (coefficients) $\lambda_j$ are in
increasing order, i.e.
\begin{equation}
\label{eqn4}
0\leqslant\lambda_1\leqslant\lambda_2\leqslant\cdots\leqslant\lambda_{D}.
\end{equation} The separable operation acting on $\ket{\psi}$ will produce an
ensemble $\{p_k,\ket{\phi_k}\}_{k=1}^N$, where
\begin{equation}
\label{eqn5} (A_k\otimes B_k)\ket{\psi}=\sqrt{p_k}\ket{\phi_k}
\end{equation} and
\begin{equation}
\label{eqn6} p_k=\matl{\psi}{A_k^\dagger A_k\otimes B_k^\dagger B_k}{\psi}.
\end{equation}

In \cite{PhysRevLett.83.1455} it was shown that such an ensemble
$\{p_k,\ket{\phi_k}\}_{k=1}^{N}$ can be produced from $\ket{\psi}$ by a suitable LOCC if
and only if the majorization inequalities
\begin{equation}
\label{eqn7} \sum_{k=1}^{N}p_kE_n(\ket{\phi_k})\leqslant E_n(\ket{\psi})
\end{equation} 
hold for $1\leqslant n\leqslant D$, where
\begin{equation}
\label{eqn8}
E_n(\ket{\psi})=\chi_n\big(\Tr_A\dyad{\psi}{\psi}\big)=\sum_{j=1}^{n}\lambda_j,
\end{equation} 
and similarly for the $\ket{\phi_k}$. Here $\Tr_A(\dyad{\psi}{\psi})$ is the
reduced density operator of $\dyad{\psi}{\psi}$ on Bob's side, and
$\chi_n(\cdot)$ is defined to be the sum of the first $n$ smallest eigenvalues
of its argument.  Note that we are assuming that $D=D_B\leq D_A$, because if
$D_B$ were greater than $D_A$ the extra zero eigenvalues in
$\Tr_A\dyad{\psi}{\psi}$ would cause confusion when using $\chi_n$.

Our main result is the following.
\begin{theorem}
  \label{thm1} The ensemble $\{p_k,\ket{\phi_k}\}_{k=1}^N$ can be produced by
  a bipartite separable operation acting on the normalized state $\ket{\psi}$
  if and only if the majorization condition defined by the collection of
  inequalities in \eqref{eqn7} is satisfied.
\end{theorem}

\begin{proof}

  To simplify the proof we assume that $D_A=D_B=D$.  If $D_A$ is larger, one
always modify each $A_k$ by following it with a suitable local unitary which
has the result that as long as the Kraus operators are acting on a fixed
$\ket{\psi}$ the action on the $A$ side takes place in a subspace of $\HC_A$
of dimension $D$.  These local unitaries do not change the Schmidt weights of
the $\ket{\phi_k}$ or alter the closure condition \eqref{eqn2}. For more
details about this ``decoupling'' see \cite{PhysRevA.76.032310}.

  When the majorization condition \eqref{eqn7} holds the result in
\cite{PhysRevLett.83.1455} guarantees the existence of an LOCC (hence
separable operation) which will produce the ensemble out of $\ket{\psi}$.  The
reverse inference, that the ensemble $\{p_k,\ket{\phi_k}\}_{k=1}^N$ defined in
\eqref{eqn5} and \eqref{eqn6} satisfies \eqref{eqn7}, follows from noting that
\begin{equation}
\label{eqn9} p_kE_n(\ket{\phi_k})=\chi_n \big(\Tr_A[A_k\otimes
B_k\dyad{\psi}{\psi}A_k^\dagger\otimes B_k^\dagger]\big),
\end{equation} 
and applying Theorem~\ref{thm2} below with $R=I_A\otimes I_B$,
corresponding to \eqref{eqn2}, 
so $\|R\|=1$.
\end{proof}

\subsection{A majorization theorem\label{sct2B}}

\begin{theorem}
\label{thm2} Let $\HC_A$ and $\HC_B$ have the same dimension $D$, let
$\ket{\psi}$ be some pure state on $\HC_A \otimes\HC_B$, and let $\{A_k\otimes
B_k\}_{k=1}^N$ be any collection of product operators on $\HC_A\otimes \HC_B$.
Then for every $1\leqslant n\leqslant D$
\begin{multline}
\label{eqn10} \sum_{k=1}^N\chi_n\big(\Tr_A[A_k\otimes
B_k\dyad{\psi}{\psi}A_k^\dagger\otimes B_k^\dagger]\big)\\
\leqslant\|R\|\chi_n\big(\Tr_A\dyad{\psi}{\psi}\big),
\end{multline} 
where $\|R\|=\sup_{\|\omega\|=1}\|R\ket{\omega}\|$ is the
largest eigenvalue of the positive operator
\begin{equation}
\label{eqn11} R=\sum_{k=1}^N A_k^\dagger A_k\otimes B_k^\dagger B_k.
\end{equation}
\end{theorem}

\begin{proof} By map-state duality
\cite{PhysRevA.76.032310,OSID.11.3,PhysRevA.73.052309}, using the Schmidt bases
of $\ket{\psi}$, we transform the state $A_k\otimes B_k\ket{\psi}$ to a map
$A_k\psi \bar B_k$, where
\begin{equation}
\label{eqn12} \psi=\sum_{j=1}^D\sqrt{\lambda_j}\dyad{a_j}{b_j}.
\end{equation} 
denotes an operator mapping $\HC_B$ to $\HC_A$, and $\bar B_k=
B_k^T$ is the transpose of $B_k$. The matrix of $\psi$ using the Schmidt bases
of $\ket{\psi}$ is diagonal, with the entries on the diagonal in
increasing order. (See Sec. II of \cite{PhysRevA.76.032310} for more details
on map-state duality.)
Upon writing the partial traces as
\begin{multline}
\label{eqn13} \Tr_A\dyad{\psi}{\psi}=\psi\psi^\dagger,\quad\\
 \Tr_A[A_k\otimes B_k\dyad{\psi}{\psi}A_k^\dagger\otimes
B_k^\dagger]= A_k\psi {\bar B_k}{\bar B_k}^\dagger\psi^\dagger A_k^\dagger,
\end{multline} 
the inequalities \eqref{eqn10} become:
\begin{equation}
\label{eqn14} \sum_{k=1}^N\chi_n(A_k \psi \bar B_k{\bar
B_k}^\dagger\psi^\dagger A_k^\dagger) \leqslant\|R\|\chi_n(\psi\psi^\dagger).
\end{equation}

For some $n$ between $1$ and $D$ write the diagonal matrix $\psi$ as
\begin{equation}
\label{eqn15} \psi=\psi_n+\tilde\psi_{n},
\end{equation} where $\psi_n$ is the same matrix but with $\lambda_{n+1},
\lambda_{n+2},\ldots$ set equal to zero, while $\tilde\psi_n$ is obtained by
setting $\lambda_1, \lambda_2,\ldots \lambda_n$ equal to zero.
Lemma~\ref{lma1}, below, tells us that for each $k$,
\begin{equation} \chi_n(A_k \psi \bar B_k {\bar B_k}^\dagger\psi^\dagger
A_k^\dagger ) \leqslant \Tr(A_k \psi_n \bar B_k {\bar
B_k}^\dagger\psi_n^\dagger A_k^\dagger ).
\label{eqn16}
\end{equation}
By map-state duality,
\begin{equation}
\label{eqn17} \Tr(A_k \psi_n \bar B_k {\bar B_k}^\dagger\psi_n^\dagger
A_k^\dagger )= \matl{\psi_n}{A_k^\dagger A_k\otimes B_k^\dagger B_k}{\psi_n}
\end{equation} where $\ket{\psi_n}$, the counterpart of $\psi_n$, is given by
\eqref{eqn3} with $D$ replaced by $n$. Inserting \eqref{eqn17} in
\eqref{eqn16} and summing over $k$, see \eqref{eqn11}, yields
\begin{eqnarray}
\label{eqn18} \sum_{k=1}^{N}&\chi_n(A_k \psi \bar B_k{\bar
B_k}^\dagger\psi^\dagger A_k^\dagger )
\leqslant\matl{\psi_n}{R}{\psi_n}\nonumber\\
&\leqslant\|R\|\ip{\psi_n}{\psi_n}=\|R\|\chi_n(\psi^\dagger\psi).
\end{eqnarray} This establishes \eqref{eqn14}, which is equivalent to
\eqref{eqn10}.
\end{proof}

\begin{lemma}
\label{lma1} Let $A$, $B$, and $\psi$ be $D\times D$ matrices, where $\psi$ is
diagonal with nonnegative diagonal elements in increasing order, and for some
$1\leqslant n\leqslant D$ let $\psi_n$ be obtained from $\psi$ by setting all
but the $n$ smallest diagonal elements equal to 0, as in \eqref{eqn15}. Then
\begin{equation}
\label{eqn19} \chi_n(A \psi B{B}^\dagger\psi^\dagger A^\dagger )\leqslant
\Tr(A \psi_n B{B}^\dagger\psi_n^\dagger A^\dagger ).
\end{equation}
\end{lemma}

\begin{proof}

The inequality
\begin{equation}
\label{eqn20} \chi_n(A \psi B{B}^\dagger\psi^\dagger A^\dagger )\leqslant
\Tr(P_nA \psi B{B}^\dagger\psi^\dagger A^\dagger P_n),
\end{equation} 
where $P_n$ is a projector (orthogonal projection operator) of rank at least
$n$, follows from the fact that for any Hermitian operator $T$ the sum of its
$n$ smallest eigenvalues is the minimum of $\Tr(P_nTP_n)$ over such $P_n$, see
page~24 of \cite{Bhatia:MatrixAnalysis}.  Choose $P_n$ to be the
projector onto the orthogonal complement of the range of $A\tilde\psi_{n}$,
where $\tilde\psi_n=\psi-\psi_n$, as in \eqref{eqn15}.  The rank of
$A\tilde\psi_{n}$ is no larger than the rank of $\tilde\psi_n$, which is
smaller than or equal to $D-n$. Thus the dimension of the range of
$A\tilde\psi_{n}$ cannot exceed $D-n$, so the rank of $P_n$ is at least $n$.
By construction, $P_nA\tilde\psi_{n}=0$, so
\begin{equation}
\label{eqn21} P_nA\psi=P_nA(\psi_n+\tilde\psi_{n})=P_nA\psi_n.
\end{equation} 
Thus with this choice of $P_n$ the right side of \eqref{eqn20} is
\begin{equation}
\label{eqn22} \Tr(P_n A \psi_n B{B}^\dagger\psi_n^\dagger A^\dagger P_n),
\end{equation} 
and this implies \eqref{eqn19}, since $P_n\leqslant I$ and $A \psi_n B
{B}^\dagger\psi_n^\dagger A^\dagger $ is positive.
\end{proof}

\section{Consequences\label{sct3}} The following are some consequences of
Theorem~\ref{thm1}.
\begin{itemize}
 \item[i)] An ensemble
$\{p_k,\ket{\phi_k}\}$ can be produced by a separable operation acting on
a bipartite entangled state
$\ket{\psi}$ if and only if it can be produced by some LOCC acting on the same
state $\ket{\psi}$.

\item[ii)] For a given bipartite $\ket{\psi}$ and separable operation
  $\{A_k\otimes B_k\}_{k=1}^N$ there is another operation of the form 
$\{\hat A_l\otimes U_l\}_{l=1}^M$, where the $U_l$ are unitary operators (and the closure condition is $\sum_{l=1}^M\hat A_l^\dagger \hat A_l=I_A$), which 
produces the same ensemble when applied to $\ket{\psi}$. Here $M$ could be
different from $N$, as two Kraus operators might yield the same $\ket{\phi_k}$. For more details about the relation between the $\{A_k,B_k\}_{k=1}^N$ set and the $\{\hat A_l\otimes U_l\}_{k=1}^M$ set see \cite{PhysRevA.63.022301}.

\item[iii)] A deterministic transformation $\ket{\psi}\rightarrow\ket{\phi}$
  by a separable operation is possible if and only if
  $E_n(\ket{\phi})\leqslant E_n(\ket{\psi})$ for every $n$ between $1$ and
  $D$, with $E_n(.)$ defined in \eqref{eqn8} This is often written as
  $\lambda_\psi\prec\lambda_\phi$, where $\lambda_\psi$ and $\lambda_\phi$ are
  vectors of the corresponding Schmidt weights.  (This extends 
  Theorem~1 in \cite{PhysRevA.76.032310}.) 

\item[iv)] The maximum probability of success for the
transformation $\ket{\psi}\rightarrow\ket{\phi}$ by a separable operation is
given by
\begin{equation}
\label{eqn23}
p_{max}^{SEP}(\ket{\psi}\rightarrow\ket{\phi})=\min_{n\in[1,D]}\frac{E_n(\ket{\psi})}{E_n(\ket{\phi})},
\end{equation} where $E_n(\cdot)$ was defined in \eqref{eqn8}.

\item[v)] An entanglement measures $E$ defined on pure bipartite
states is nonincreasing on average under separable
operations, which is to say
\begin{equation}
\label{eqn24} E(\ket{\psi})\geqslant\sum_{k=1}^N p_kE(\ket{\phi_k})
\end{equation}
if and only if it is similarly nonincreasing under LOCC.

\item[vi)] Let
\begin{equation}
\label{eqn25} \hat E(\rho)=\inf\sum_{i}p_iE(\ket{\psi_i}),
\end{equation} 
with the infimum over all ensembles $\{p_i,\ket{\psi_i}\}$ yielding the
density operator $\rho=\sum_ip_i\dyad{\psi_i}{\psi_i}$, be the convex roof
extension of a pure state entanglement measure $E$ that is monotone on pure
states in the sense of \eqref{eqn24}.  Then $\hat E$ is monotone on mixed
states in the sense that
\begin{equation}
\label{eqn26} \hat E(\rho)\geqslant\sum_{k=1}^N p_k\hat E(\sigma_k)
\end{equation} 
for any ensemble $\{p_k,\sigma_k\}$ produced from $\rho$ by separable
operations.

\end{itemize}

The result (i) is an immediate consequence of Theorem~\ref{thm1}, as the
same majorization condition applies for both separable and LOCC.  Then
(ii), (iii), and (iv) are immediate consequences of known results, in 
 \cite{PhysRevA.63.022301}, \cite{PhysRevLett.83.436}, and 
\cite{PhysRevLett.83.1046}, respectively, for LOCC.  The result (v) is an
obvious consequence of (i), whereas (vi) follows from general arguments
about convex roof extensions; see Sec.~XV.C.2
of \cite{quantph.0702225}.

\section{Conclusion\label{sct4}} 

We have shown that possible ensembles of states produced by applying a separable
operation to a bipartite entangled pure state can be exactly characterized
through a majorization condition, the collection of inequalities \eqref{eqn7} 
for different $n$.  These have long been known to be necessary and sufficient
conditions for producing such an ensemble using LOCC, so their extension to
the full class of separable operations is not altogether surprising, even if
our proof is not altogether straightforward.  

Connecting the full set of separable operations with the more specialized LOCC
class immediately yields several significant consequences for the former, as
indicated in the list in Sec.~\ref{sct3}, because much is already known about
the latter.  Of particular significance is that various entanglement measures
are monotone, meaning they cannot increase, under separable
operations---something expected on intuitive grounds, but now rigorously
proved.  Since such monotonicity under LOCC has long been considered a
necessary, or at least a very desirable condition for any ``reasonable''
entanglement measure on mixed states (see Sec.~XV.B of \cite{quantph.0702225}), one
wonders whether monotonicity under separable operations, in principle a
stronger condition, might be an equally good or even superior desideratum.

Our results apply only to bipartite states, but separable operations and the
LOCC subclass can both be defined for multipartite systems.  Might it be that
in the multipartite case the ensemble produced by applying a separable
operation to a pure entangled state could also be produced by some LOCC
applied to the same state?  It might be, but proving it would require very
different methods than used here.  There are no simple multipartite analogs of
the Schmidt representation \eqref{eqn3}, the majorization condition
\eqref{eqn7}, or map-state duality.

Even in the bipartite case we still know very little about separable
operations which are \emph{not} LOCC, aside from the fact that they exist and
can be used to distinguish certain collections of orthogonal states more
effectively than LOCC.  The results in this paper contribute only indirectly
to a better understanding of this matter: looking at what a separable
operation does when applied to a \emph{single} entangled state will not help;
one must ask what it does to several different states.

\begin{acknowledgments} We thank Li Yu for useful comments. The research
described here received support from the National Science Foundation through
Grant No. PHY-0456951.

\end{acknowledgments}


\end{document}